\documentclass[doublecol]{epl2}
\usepackage{graphicx}
\usepackage{amsmath}
\usepackage{amssymb}

\newcommand{\comment}[1]{}
\newcommand{\BEQ}{\begin{equation}}
\newcommand{\EEQ}{\end{equation}}
\newcommand{\BEA}{\begin{eqnarray}}
\newcommand{\EEA}{\end{eqnarray}}
\newcommand{\bea}{\begin{eqnarray}}
\newcommand{\eea}{\end{eqnarray}}

\renewcommand{\d}{{\rm d}}

\newcommand{\e}{{\bf e}}
\newcommand{\0}{{\bf 0}}

\title{Work extraction from microcanonical bath}

\author{Armen E. Allahverdyan \inst{1,2} and Karen V. Hovhannisyan \inst{1}}
\shortauthor{Allahverdyan and Hovhannisyan}

\institute{ \inst{1} Yerevan Physics Institute, Alikhanian Brothers Street 2, Yerevan 375036, Armenia \\
\inst{2} Laboratoire de Physique Statistique et Syst\`emes Complexes,
ISMANS, 44 ave. Bartholdi, 72000 Le Mans, France
}

\abstract{ We determine the maximal work extractable via a cyclic
Hamiltonian process from a positive-temperature ($T>0$) microcanonical
state of a $N\gg 1$ spin bath. The work is much smaller than the
total energy of the bath, but can be still much larger than the energy
of a single bath spin, e.g. it can scale as ${\cal O}(\sqrt{N\ln
N})$. Qualitatively same results are obtained for those cases, where the
canonical state is unstable (e.g., due to a negative specific heat) and
the microcanonical state is the only description of equilibrium. For a
system coupled to a microcanonical bath the concept of free energy does
{\it not generally} apply, since such a system|starting from the canonical
equilibrium density matrix $\rho_T$ at the bath temperature $T$|can
enhance the work extracted from the microcanonical bath without changing
its state $\rho_T$. This is impossible for any system coupled to a
canonical thermal bath due to the relation between the maximal work and
free energy. But the concept of free energy still applies for a sufficiently large
$T$. Here we find a compact expression for the {\it microcanonical
free-energy} and show that in contrast to the canonical case it contains a
{\it linear entropy} instead of the von Neumann entropy.}


\begin{document}

\maketitle

How much work can be extracted from a state of a physical system via
cyclic processes? This question governs our understanding of energy
conversion and storage, and hence is central for
thermodynamics \cite{landau,passivity,gyfto,abn,jan,vedral,passo,q_t}.
The basic answer, known as the Thomson's formulation of the second law,
is that an equilibrium state cannot yield work. This formulation is an
axiom in thermodynamics, but its first-principle derivations were given
in literature for a {\it canonical} (Gibbsian) equilibrium state
\cite{passivity}. The main consequence of the Thomson's formulation is
that only non-equilibrium states can be sources of work. The maximal
work extractable from such states via a cyclic process was
studied both for macroscopic \cite{landau} and finite
systems \cite{gyfto,abn,jan}.

One instance of the maximal work is especially well-known, because it
provides the physical meaning of free energy \cite{landau}. Consider a
quantum system with Hamiltonian $H$ and initial density matrix $\rho$.
This system is in contact with a
canonical thermal bath at temperature $T$. External fields act cyclically
on the system + bath. Assuming no system-bath coupling
both initially and finally, the maximal work extracted by the fields
reads \cite{landau,f0}
\BEA
W_{\rm max}= F[\rho]-F[\rho_{\rm eq}], ~~~ F[\rho]={\rm tr}(\rho H) +T{\rm tr}(\rho\ln \rho),
\label{baratino}
\EEA
where $F[\rho]$ is the free energy and $\rho_{\rm eq}=e^{-H/T}/{\rm
tr}[e^{-H/T}]$ is the canonical equilibrium state of the system, which is
its final state after work-extraction \cite{landau}. The maximal work
is determined by the deviation of $\rho$ from its canonical equilibrium
value $\rho_{\rm eq}$ as quantified by the free energy (\ref{baratino}).

One notes however that {\it all} above results refer to a specific
notion of equilibrium, viz. the canonical state. Another concept of
equilibrium is given by the microcanonical state, which describes an
isolated system equilibrated due to its internal mechanism \cite{polko},
or an open system coupled weakly to its environment (so weak that no
energy is exchanged) \cite{q_t,hove}. This is a more fundamental notion
of equilibrium: {\it (i)} under certain conditions the canonical state can
be derived from it for a weakly coupled subsystem \cite{landau}. {\it
(ii)} In contrast to the canonical state, whose preparation refers to an
external thermal bath, the microcanonical state can be applied to a closed
few-body system provided that it satisfies certain chaoticity features
\cite{berdi,borgo}. {\it (iii)} Since {\it local} stability conditions
of the canonical state are more demanding|a fact closely related to the
no work-extraction feature\cite{landau}|there are situations,
where the equilibrium can be described by the microcanonical state only,
since the canonical state for them is unstable \cite{longrange}. For such
systems, frequently realized via long-range interactions, the entropy is
a non-concave function of energy, and hence the notorious {\it
macroscopic} equivalence between canonical and microcanonical state is
broken \cite{longrange}. Even if this macroscopic equivalence holds, it
is by no means obvious that in the argument around (\ref{baratino}) one
can substitute the canonical state of the thermal bath by the microcanonical
state [with the same temperature], because {\it in general} the work
(\ref{baratino}) is not a macroscopic quantity, i.e. it does not scale
with the number of bath particles. This is however widely done in
literature, e.g., when introducing the free energy as in
(\ref{baratino}) one basically never specifies the equilibrium state of
the bath; see, e.g. \cite{landau,vedral}.

We revisit the maximal work-extraction problem for a thermal
bath in a quantum microcanonical state. It was noted already some work can
be extracted via a cyclic Hamiltonian process from a {\it few-particle}
microcanonical system \cite{passo}. Recent papers studied
to which extent the extraction of work from one-particle classical
microcanonical system can be carried out by physically realistic
Hamiltonians \cite{parrondo,jar}. Our purposes here are different:

$\bullet$ We focus on finding the maximal amount of work extractable from a
{\it macroscopic} microcanonical state of $N\gg 1$ particle thermal bath.

$\bullet$ We also determine the work extracted via a system coupled to a
microcanonical thermal bath, and check whether the reasoning
(\ref{baratino}) generalizes {\it at least} qualitatively, i.e. whether
the concept of free energy applies to the microcanonical situation.

The subject of work extraction via a system coupled to a thermal bath is
an active research topic.  Refs.~\cite{jar,armen_epl,seifert,theo,ueda}
discuss various set-ups for this problem: quantum, classical, with or
without state-dependent feedback {\it etc}.  Recall that
(\ref{baratino}) is at the core of relations between statistical
thermodynamics and information theory \cite{vedral}.  The term ${\rm
tr}([\rho-\rho_{\rm eq}]H)$ in the right-hand-side of (\ref{baratino})
is the energy extracted from the system, while the remaining (entropic)
part comes from the bath. If ${\rm tr}([\rho-\rho_{\rm eq}]H)$ is
negligible (e.g., because $H$ contains only few almost degenerate energy
levels), the work is extracted from the bath due to the difference
between the initial entropy and its canonical equilibrium value. This
relation between the entropy and work is the essential part of
information driven engines (e.g., Szilard's engine) \cite{vedral}. In
contrast, various forms of fuel operate due to the initial
non-equilibrium energy, i.e. the term ${\rm tr}([\rho-\rho_{\rm eq}]H)$
in (\ref{baratino}).

\comment{{\bf Organization of the paper.} We continue by recalling the basic
model of the microcanonical bath and its main features. Then we determine
the maximal work extractable from it in two regimes: the bath
temperature comparable with the single particle energy and much larger
than that energy. We also determine how the extracted work depends on
the width of the microcanonical state. Finally we study the work
extractable via a smaller system coupled to the microcanonical bath. }

{\bf Microcanonical thermal bath.} The microcanonical state is characterized by two parameters: energy $E$ and width $\sigma$ \cite{landau}.
The corresponding density matrix is diagonal in the energy representation, all energies within the interval $[E,E-\sigma]$
have equal probabilities, all other energies have probability zero. For a $N\gg 1$-particle system the number $d(E,\sigma)$
of energy levels within the interval $[E,E-\sigma]$ defines the microcanonical entropy \cite{landau}:
\BEA
\label{entropy}
S(E) = \ln d(E,\sigma)={\cal O}(N),
\EEA
where the choice of $\sigma$ should not influence the leading ${\cal
O}(N)$ behavior of $S(E)$. Eq.~(\ref{entropy}) is the von Neumann
entropy for the microcanonical density matrix (\ref{ino}); see also
\cite{f1}. For clarity we want to work with a specific model of a
macroscopic microcanonical system (bath). This is the basic model of the
field: $N\gg 1$ uncoupled two-level spins; each spin has energies $0$
and $\delta>0$ \cite{q_t}. Some of our results extend to more general
bath models, as seen below.

The bath Hamiltonian reads (${\rm
diag}[\ldots]$ means diagonal matrix in the energy representation)
\BEA
\label{hamo}
H={\rm diag}[0,\, \delta, \, 2\delta\ldots, \, \delta N],
\EEA
where each element $\delta k$ is repeated $d_k$ times,
\BEA
d_k\equiv \frac{N!}{k!(N-k)!}.
\label{kaban}
\EEA
Hence every energy shell $\delta k$ is $d_k$--degenerate.
Denote
\BEA
\label{khorasan}
\e_k=(1,\ldots,1), ~~~\0_k=(0,\ldots,0), ~~ D_k\equiv{\sum}_{m=0}^{k}d_m,
\EEA
where $\e_k$ ($\0_k$) is the vector of $k$ $1$'s ($0$'s).

For the present model of bath the microcanonical state is easy to define:
all energies $\delta M$ have equal probability $\frac{1}{d_M}$; all
other energies have zero probability. Thus we put $\sigma\to 0$, the
{\it minimal} thermodynamically consistent width for this model.  Note
that the degenerace of the energy levels is convenient, since it allows
to set $\sigma\to 0$.  It is however not essential: an effective
degeneracy will be anyhow regained for a small but finite $\sigma>0$,
since the energy levels of a macroscopic system are located very densely
\cite{landau}.
The bath initial state reads in representation (\ref{hamo})
\BEA
\label{ino}
\Omega_{\rm i}=\frac{1}{d_M}{\rm diag}[\,\0_{D_{M-1}}, \e_{d_M},\0_{D_N-D_{M}}\,].
\EEA
For $N\gg 1$ this microcanonical state does have desired features
expected from thermodynamics, e.g. macroscopic equivalence
with the canonical state, equilibration of a small subsystem,
third law \cite{q_t}. The density matrix of a single bath spin is
Gibbsian $\propto {\rm diag}[1,e^{-\delta/T}]$ with \cite{q_t}
\BEA
e^{-\delta/T}=m/(1-m), ~~ m=M/N.
\label{chermo}
\EEA
The  same $T$ is recovered as microcanonical temperature \cite{landau}
\BEA
\label{po}
{1}/{T}={\partial S(E) }/{\partial E}, ~~~ E= \delta M, ~~~ S(E)=\ln d_M,
\EEA
where $S(E)$ is the microcanonical entropy (\ref{entropy}). This equivalence
can be shown via formula (\ref{koza}) that is proven below.

Note that although the spins are uncoupled, the microcanonical state does
not reduce to the tensor product of the separate spin states (otherwise
it would amount to the canonical state). It contains inter-spin
correlations. Ultimately, this is the reason why, as seen below, a microcanonical bath
can yield work in a cyclic process.

We restrict ourselves with $M/N\leq 1/2$, i.e. {\it positive}
temperatures. The case with $M/N> 1/2$ is definitely less
interesting, because now each spin of the bath is in a state
with a negative temperature. Such states are trivially active,
i.e. they yield work in a cyclic process.

{\bf Work extraction.} At some initial time $t=0$ the bath Hamiltonian
$H(t)$ becomes time-dependent due to interaction with sources of work.
Consider a cyclic process
\BEA
\label{dumbadze}
H(0)=H(\tau)=H,
\EEA
where $\tau$ is the final time.
The work extracted in this thermally isolated cyclic Hamiltonian process is
\BEA
\label{beau}
W={\rm tr}(H[\Omega_{\rm i}-\Omega_{\rm f}])=\delta M-{\rm tr}(H\Omega_{\rm f}), \,\, \Omega_{\rm f}=U\Omega_{\rm i}U^\dagger,\,
\EEA
where $\Omega_{\rm f}$ is the final state of the bath, and $U={\cal
T}e^{-(i/\hbar)\int_0^\tau \d s H(s)}$;
${\cal T}$ means chronologization. Conversely, for a given
unitary $U$ one can construct a class of Hamiltonians that generate $U$
and satisfies (\ref{dumbadze}) \cite{abn}.

Condition (\ref{dumbadze}) is necessary for the system to be an autonomous
carrier of energy that should deliver work to another system (e.g., to a
work-source) via an interaction which switches on and off at
well-defined times. Hence this is a cyclic Hamiltonian process.

We now maximize the work $W$|or minimize the final energy ${\rm tr}(H\Omega_{\rm f})$|over
all cyclic Hamiltonians, i.e. over unitary operators $U$. Note from (\ref{beau}) that
\BEA
\label{khan1}
{\rm tr}(H\Omega_{\rm f})={\sum}_{a,b=1}^{2^N}E_a C_{ab}
\langle b| \Omega_{\rm i} |b\rangle ,~~ C_{ab} \equiv |\langle b| U |a\rangle|^2,\\
C_{ab}  \geq 0, ~~ {\sum}_{a=1}^{2^N} C_{ab} = {\sum}_{b=1}^{2^N} C_{ab} =1,
\label{khan2}
\EEA
where $\{E_a\}_{a=1}^{2^N}$ and $\{|a\rangle\}_{a=1}^{2^N}$ are,
respecitively, the eigenvalues and eigenvectors of $H$ [see
(\ref{hamo})], and the elements $\{\langle b| \Omega_{\rm i}
|b\rangle\}_{b=1}^{2^N}$ are defined in (\ref{ino}). Three conditions in
(\ref{khan2}) mean that the matrix $C_{ab}$ is double-stochastic \cite{olkin}.
Conversely, every such matrix can be represented as $C_{ab} \equiv
|\langle b| U |a\rangle|^2$ for some unitary $U$ \cite{olkin}. Every
double-stochastic matrix equals to a convex sum of permutation matrices
$\Pi^{[\alpha]}$ (Birkhoff's theorem \cite{olkin}):
$C={\sum}_\alpha \lambda_\alpha \Pi^{[\alpha]}$, ${\sum}_\alpha\lambda_\alpha=1$,
$\lambda_\alpha\geq 0$,
where each matrix $\Pi^{[\alpha]}$ acting on a column-vector $x$ amounts
to permuting (in a certain way) the elements of $x$. Eq.~(\ref{khan1})
shows that ${\rm tr}(H\Omega_{\rm f})$ is a linear function of the
matrix $C=\{C_{ab}\}$. Hence its minimum over the unitary operators $U$,
that is its minimum over double-stochastic matrices $C_{ab}$, is reached
for $C_{ab}$ equal to some permutation matrix $\widetilde{\Pi}$. It is
clear from (\ref{khan1}) that $\widetilde{\Pi}$, when acting on the
vector $\langle b| \Omega_{\rm i} |b\rangle$ permutes its elements such
that all its non-zero [equal to each other] elements concentrate at
lowest energies $\{E_a\}_{a=1}^{2^N}$ \cite{abn}. For the final state we
have $\langle a| \Omega_{\rm f}
|a\rangle={\sum}_{b}\widetilde{\Pi}_{ab}\langle b| \Omega_{\rm i}
|b\rangle$.  Hence the lowest-energy final state compatible with
$\Omega_{\rm i}$ reads
\BEA
\label{fino}
\Omega_{\rm f}=\frac{1}{d_M}{\rm diag}[\,\e_{d_M}, \0_{D_N-d_{M}}\,].
\EEA
Once $\widetilde{\Pi}$ is found we can employ the standard procedure of
constructing the corresponding unitary operator $U$ and the cyclic
Hamiltonian \cite{abn}.
Note that $\widetilde{\Pi}$ does depend on the energy of the initial
state $\Omega_{\rm i}$: $\widetilde{\Pi}$
applied on a microcanonical state with a different energy will not lead to
the maximal work-extraction. This does not differ from (say)
the ordinary Carnot cycle, whose implementation also demands knowing the
initial state of the working body.

The maximal work $W_{\rm max}$ reads from (\ref{fino}, \ref{beau}): $W_{\rm max}= \delta M-\frac{\delta}{d_M}[
{\sum}_{k=0}^{M-\ell} \, k \, d_k +(M-\ell+1)(d_M-{\sum}_{k=0}^{M-\ell}d_k )]$. After summation by parts,
\BEA
\label{s}
W_{\rm max}= \delta\left[
\ell-1+\frac{1}{d_M}{\sum}_{k=0}^{M-\ell}D_k
\right],
\EEA
where $D_k$ is defined in (\ref{khorasan}), and where integer $\ell=\ell(M)$ is found from
\BEA
\label{simone}
{\sum}_{k=0}^{M-\ell+1}\, d_k > d_M\geq {\sum}_{k=0}^{M-\ell} \, d_k.
\EEA
Eq.~(\ref{s}, \ref{simone}) hold for any microcanonical state (\ref{hamo}, \ref{ino});
the specific form (\ref{kaban}) is not necessary.

We shall now calculate $W_{\rm max}$ for two limits: $T\to\infty$ and a finite $N$, and then $N\to\infty$ and a finite $T$.

{\bf Doubly maximized work.} We set the number of spins $N$ to a large,
but a finite number, and maximize $W_{\rm max}(T)$ over all positive
temperatures of the $N$-spin bath. The maximum is reached for $T=\infty$
(or $M= N/2$ as (\ref{chermo}) shows) and provides an upper bound for
the work extractable from the positive temperature bath. We now
calculate $W_{\rm max}(\infty)$. Consider the sum${\sum}_{k=0}^{\frac{N}{2}-\ell}d_k={\sum}_{m=\ell}^{\frac{N}{2}}d_{\frac{N}{2}-m}$
in (\ref{simone}). The dominant summation region is $m\sim \ell$. We
shall see below that $\ell\ll N$.  Hence for (\ref{kaban}) we use the
Gaussian approximation $d_{\frac{N}{2}-m}=2^{\frac{N}{2}}
e^{-\frac{m^2}{N/2}}$ \cite{f2}, change the sum to integral, and find
$\ell$ from $1= \int_{\ell}^\infty d x \,e^{-\frac{x^2}{(N/2)}}$: $\ell
= \sqrt{\frac{N}{4}\ln\left[ \frac{N}{4\ln (N/4)} \right] +
\frac{N}{4}{\cal O} \left[ \frac{\ln \ln (N/4)}{\ln (N/4)} \right]}$.
The second term under square root is negligible if $\ln N$ is large.
Eq.~(\ref{s}) then implies (for $N\gg 1$):
\BEA
W_{\rm max} =\delta\int_{\ell}^\infty\d x\,x e^{-\frac{x^2}{(N/2)}}
=\delta\ell\approx\frac{\delta}{2}\sqrt{N\ln N}.
\label{grant}
\EEA
This is a reachable upper bound for the work extractable from the $N$-spin microcanonical bath.

{\bf Finite temperatures and thermodynamic limit.}
We employ (\ref{kaban}) and assume the standard thermodynamic limit: $m=M/N<1/2$
(and hence $T>0$) in (\ref{chermo}) is a fixed finite number for
$M,N\to\infty$.  We note from (\ref{kaban}) and (\ref{chermo}) that for any fixed finite
numbers $m$, $\ell$ and $N\to\infty$,
\BEA
\frac{d_{Nm-\ell}}{d_{Nm}}= e^{-\ell\delta/T}\, [\,1+{\cal O}(\frac{1}{N})\,].
\label{koza}
\EEA
Since the sums in (\ref{s}, \ref{simone}) are dominated by their largest terms, using
(\ref{koza}) (with $m<1$ and integer $\ell$) amounts to calculating these sums via geometrical progression, e.g.,
$\frac{1}{d_M}{\sum}_{k=0}^{M-\ell+1}\, d_k={\sum}_{k=\ell-1}^\infty \, e^{-\delta k/T}$.
We get for $\ell$ and $W_{\rm max}$
\BEA
\label{brabant}
&& W_{\rm max}(T)=\delta\left[ \ell-1+v^{\ell}~(1-v)^{-2} \right],\\
\label{brab}
&& v\equiv e^{-\delta/T}, \qquad \ell=\left\lceil \frac{\ln(1-v)}{\ln v}\right\rceil,
\EEA
where $\lceil x \rceil$ is the ceiling (upper) integer part of $x$,
e.g., $\lceil 0.99\rceil=1$, $\lceil -0.99\rceil=0$.  According to
(\ref{brab}), $\ell$ grows to infinity with $T$: $\ell=1$ for
$e^{-\delta/T}\leq \frac{1}{2}$, $\ell=2$ for $\frac{2}{1+\sqrt{5}}\geq
e^{-\delta/T}\geq \frac{1}{2}$, $\ell=3$ for $0.68232\geq
e^{-\delta/T}\geq \frac{2}{1+\sqrt{5}}$ {\it etc}.

\begin{figure}
\includegraphics[width=6.5cm]{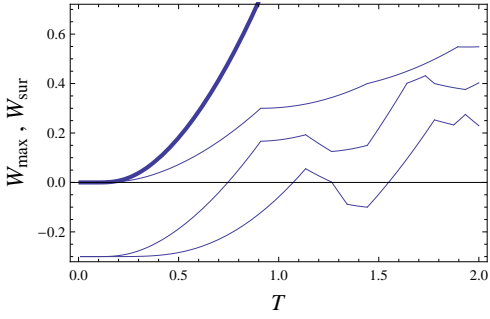}
\caption{Thick curve: $W_{\rm max}$ given by (\ref{s}) as a function of temperature $T$ for $\delta=1$.
Normal curves show $W_{\rm sur}$ given by (\ref{barak})
as a function of $T$ for $\delta=1$, $\pi_1=0.3$
and (from top to bottom): $\alpha=\lfloor \frac{\epsilon}{\delta}\rfloor=0,\,1,\,2$.
}
\label{fig1}
\end{figure}

$W_{\rm max}(T)$ is a continuous function of $T$, but $\frac{\d W_{\rm max}}{\d
T}$ has jumps at the temperatures, where $\ell$ changes, and one energy shell in the final
density matrix (\ref{fino}) is completely filled; see
(\ref{simone}) and Fig.~\ref{fig1}. Hence
\BEA
\label{ort}
W_{\rm max}(T)={\cal O}(\delta) ~~{\rm for}~~ N\gg 1 ~~{\rm and}~~ T={\cal O}(\delta).
\EEA
The situation is symmetric with respect to different spins of the bath.
Hence after the work-extraction the initial energy of each bath spin
changes negligibly $={\cal O}(\frac{1}{N})$. The final state of each
spin is diagonal in the energy representation and thus after
work-extraction it has a well-defined temperature that differs from the
initial temperature by ${\cal O}(\frac{1}{N})$. Recall that the
Gibbsian state as such cannot yield work in a cyclic process
(\ref{dumbadze}) \cite{passivity}. Hence the work is extracted due to
inter-spin correlations present initially in the microcanonical state.

Eq.~(\ref{brabant}) shows that $W_{\rm max}(T)$ increases faster than
$T$:
\BEA
\label{gogor}
W_{\rm max}(T)=T\left[ \,\ln\left(\frac{T}{\delta}  \right)+1\,\right]~~{\rm for}~~T\gg \delta,
\EEA
where we used $\left\lceil \frac{\ln(1-v)}{\ln v}\right\rceil\approx \frac{\ln(1-v)}{\ln v}$.
Eq.~(\ref{gogor}) is practically good already for $T>1.5\,\delta$.
Now $W_{\rm max}$ can be much larger than the energy of a single bath spin. In the limit
$T\gg \delta$ this energy is equal to $\delta/2$; see (\ref{chermo}).

{\bf Microcanonical states not equivalent to the canonical
one.} Eq.~(\ref{ort}) does not depend on the concrete form (\ref{kaban})
of $d_k$. What is needed for (\ref{ort}) is that the sum
${\sum}_{k=0}^Md_k$ is dominated by its last term $d_M$. Then
(\ref{simone}) implies $\ell={\cal O}(1)$, and (\ref{s}) leads to
(\ref{ort}). Hence (\ref{ort}) generalizes the Thomson's formulation of
the second law to the microcanonical situation. In particular, (\ref{ort}) holds
for those $d_k$, where the macroscopic equivalence between microcanonical and
canonical states is violated. As an example consider (\ref{hamo}, \ref{ino}) with $d_M=e^{N(M/N)^2}$.
This spectrum satisfies all above conditions and leads to (\ref{ort}).
Now the entropy $\ln d_M$ is a {\it convex function} of energy $\delta
M$. Hence the specific heat $C=[\frac{\d T}{\d (\delta M)}]^{-1}$
calculated from (\ref{po}) is {\it negative}, and the macroscopic
equivalence between canonical and microcanonical states is clearly violated,
because $C>0$ is an automatic consequence
of the canonical state \cite{landau}. Such convex-entropy spectra are
realized in macroscopic long-range interacting systems \cite{longrange}.

\comment{
Recall again that $C>0$ relates to the local (small perturbations)
stability of the canonical state, but it need not
hold for a locally stable microcanonical state \cite{landau,longrange}. }

Another example of convex entropy and canonical-microcanonical
non-equivalence, where still (\ref{ort}) holds, is the {\it first-order
microcanonical} phase transition \cite{landau,longrange}, where in the
vicinity of some critical energy $E_c$, $\frac{\d S(E)}{\d E}$ has a
jump: $\frac{\d S(E)}{\d E}|_{E\to E_c+}=\frac{1}{T_h}$, $\frac{\d
S(E)}{\d E}|_{E\to E_c-}=\frac{1}{T_l}$.  This describes coexistence of
two phases with different temperatures.  Since a more stable phase
should have a larger entropy, we get $T_h<T_l$ \cite{landau}.  The above
non-equivalence is seen here, because at a canonical
first-order phase transition different phases have the same temperature
\cite{landau}. Even though two phases at different temperatures do
co-exist, the extracted work has the same order of magnitude (\ref{ort}) as for a
homogeneous-temperature microcanonical state.

\comment{
These examples confirm that the generalization of the
Thomson's formulation is non-trivial.}

\comment{
Let us show that the rough scaling $W_{\rm max}\sim\sqrt{N}$ does not
depend on details of the considered model. It holds for any microcanonical
state, where the entropy $S(E)=Ns(\varepsilon)$ [$\varepsilon=E/N$]
given by (\ref{po}) has a maximum at some $\varepsilon\to
\varepsilon_0$:
\BEA
\label{balasan}
s(\varepsilon)-s(\varepsilon_0)={\sum}_{k\geq 2}a_k(\varepsilon-\varepsilon_0)^k.
\EEA
In the considered model and for $\varepsilon\approx \varepsilon_0$ we
can retain in (\ref{balasan}) only the term
$a_2(\varepsilon-\varepsilon_0)^2$. Hence $d_E$ is approximately
constant for $E-E_0\sim \sqrt{N}$ and (\ref{s}) produces $W_{\rm
max}\sim \sqrt{N}$. Imagine a model spectrum such that $a_2$ in (\ref{balasan})
vanishes: $s(\varepsilon)$ bends at $\varepsilon_0$.
We lack a good physical realization of such a spectrum, but
its existence is not forbidden. Now $d_E$ is constant for $E-E_0\sim
N^{2/3}$, hence $W_{\rm max}\sim N^{2/3}$. Thus the scaling of the
maximal work with the number of particles depends on the behaviour of
$d_E$ for very large temperature; the scaling $W_{\rm max}\sim \sqrt{N}$
is rather general, but not exclusive. If in
(\ref{balasan}) the first-order term $a_1(\varepsilon-\varepsilon_0)$
does not vanish, we revert to (\ref{ort}). }

\comment{
{\bf Microcanonical state with a larger width.} So far we defined the
microcanonical state with the minimal thermodynamically consistent width
$\sigma\to 0$; see our discussion around (\ref{entropy}).  Expectedly,
the maximal work will decay for a larger width, e.g., for a finite
temperature [regime (\ref{koza})] we have $W_{\rm max}\ll \delta$ if
$e^{-\delta\sigma/T}\ll 1$.

The high-temperature limit $M\to\frac{N}{2}$ is more interesting. Assume
that $\sigma=\widetilde{\sigma}\sqrt{N}$ with a finite
$\widetilde{\sigma}$. This scale of the width extends the equivalence
between canonical and microcanonical state to fluctuations of energy
\cite{landau}.  Adapting the derivation of (\ref{grant}) one gets
asymptotically
\BEA
\label{maxim}
W_{\rm max}=\delta\sqrt{\frac{N}{2}}f(\widetilde{\sigma}),\,\,\,\,\,\,
f(\widetilde{\sigma})= \frac{e^{-\widetilde{\sigma}^2}+e^{-\widetilde{\ell}^2}-1}
{2\int_0^{\widetilde{\sigma}} \d x\, e^{-x^2} },
\EEA
where $\widetilde{\ell}=\frac{\ell}{\sqrt{N}}$ is found from
$\int_0^{\widetilde{\sigma}}\d x\, e^{-x^2}
=\int_{\widetilde{\ell}}^\infty\d x \,e^{-x^2}$. $f(\widetilde{\sigma})$
grows as $\sqrt{-\ln \widetilde{\sigma}}$ for $\widetilde{\sigma}\to 0$
indicating the regime change to $W_{\rm max}={\cal O}(\sqrt{N\ln N})$.
For $\widetilde{\sigma}\gg 1$, $f(\widetilde{\sigma})$ tends to zero as
$e^{-\widetilde{\sigma}^2}$. But for a finite $\widetilde{\sigma}$ it is
positive and finite, e.g. $f(3)\approx 0.024$, showing that changing
$\sigma$ from $0$ to $\propto\sqrt{N}$ decreases $W_{\rm max}$ from
${\cal O}(\sqrt{N\ln N})$ to ${\cal O}(\sqrt{N})$.

Hence in the high-temperature regime
the maximal work is persistent with respect
to increasing $\sigma$.}

{\bf System coupled to the bath.} An important instance of the maximal
work problem is the amount of work extractable from a thermal bath in
the presence of a smaller system coupled to it; see (\ref{baratino}).
How much work can be extracted from a combined state of a two level
system with energies $0$ and $\epsilon>0$ and the microcanonical thermal
bath? Answering this question will alow us to understand to which extent
the concept of the free energy applies to the microcanonical situation.
Before starting the analysis we should like to stress again that so far the
statistical physics literature does not distinguish between canonical and
microcanonical situations when introducing and applying the free energy
concept; see e.g. \cite{landau}.

Let the initial density matrix of the two-level system be $\rho_{\rm
i}$; its eigenvalus are $\pi_0>\pi_1$. The spectrum of the overall
initial state $R_{\rm i}\equiv\rho_{\rm i}\otimes\Omega_{\rm i}$ reads
[see (\ref{ino}, \ref{khorasan})]
\BEA
\label{manu}
{\rm Spec}[R_{\rm i}]= \frac{1}{d_M}
[\, \pi_0\e_{d_M},\pi_1\e_{d_M},\0_{2D_N-2d_{M}} \,],~ \pi_0>\pi_1.
\EEA
Both initially and finally the two-level system and bath do not interact. Hence
the overall Hamiltonian ${\cal H}$ reads
\BEA
\label{brut}
{\cal H}&=&{\cal H}(\tau)= H_{\rm S}\otimes 1+1\otimes H\\
&=&{\rm diag}[0,\, \delta, \, 2\delta\ldots, \, \delta N, \,\epsilon,\, \epsilon+\delta,\ldots, \epsilon+\delta N],~~
\label{optimates}
\EEA
where $H$ is given by (\ref{hamo}), and $H_{\rm S}$ is the two-level
Hamiltonian with energies $0$ and $\epsilon$.  We recall that each
symbol $k\delta$ (or $\epsilon+k\delta$) in (\ref{optimates}) is
repeated $d_k$ times. Once we consider unitary work-extraction
processes, the final state of the overall system will have the same
eigenvalues (\ref{manu}). Recalling our discussion between (\ref{khan1})
and (\ref{fino}) it should be clear that the minimal final energy for
the overall system is achieved for the unitary operator that forces $R_{\rm f}$ to
have the same eigenvectors as ${\cal H}$ and permutes the eigenvalues
(\ref{manu}) such that the largest eigenvalue is matched with the
smallest energy, next to the largest eigenvalue with the next to the
smallest energy and so on. Note that ${\rm Spec}[R_{\rm i}]$ is already
ordered in a non-increasing way. It remains to order (\ref{optimates})
in a non-decreasing way and write the lowest final average energy as
scalar product of two vectors
\begin{gather}
\label{oro}
{\rm tr}({\cal H}R_{\rm f}) =
{\rm Spec}[R_{\rm i}]\,\cdot\,
[\,0,\,\delta,\ldots,\alpha\delta, \,\epsilon, \\
(\alpha+1)\delta,\, \epsilon+\delta,\ldots,\,
\delta N,\, \epsilon+\delta (N-\alpha),\ldots, \epsilon+\delta N\,],
\nonumber
\end{gather}
where $\alpha=\lfloor \frac{\epsilon}{\delta}\rfloor$, and $\lfloor x \rfloor$
is the floor (lower) integer part of $x$, e.g., $\lfloor 0.99\rfloor=0$, $\lfloor -0.99\rfloor=-1$.

Consider the work extracted from the overall system that is maximized
over all unitary dynamic operators. Since the system and bath do not
interact both initially and finally, this work separates into two
parts coming, respectively, from the system and bath
[see (\ref{beau}, \ref{s})]:
\BEA
\label{gal}
{\rm tr}({\cal H}[R_{\rm i}-R_{\rm f}]) =W_{\rm max}+W_{\rm sur}+{\rm tr}(H_{\rm S}[\rho_{\rm i}-\rho_{\rm f}]),
\EEA
where ${\rm tr}(H_{\rm S}[\rho_{\rm i}-\rho_{\rm f}])$ is the energy
change of the two-level system. $W_{\rm max}+W_{\rm sur}$ is the work coming from the bath. Here
$W_{\rm max}$ is given by (\ref{s}) (the maximal work extracted from the bath alone) and we defined
the surplus work $W_{\rm sur}$ (work extracted from the bath, but due to the system).

Obviously, $W_{\rm sur}+{\rm tr}(H_{\rm S}[\rho_{\rm i}-\rho_{\rm
f}])\geq 0$, since ${\rm tr}({\cal H}[R_{\rm i}-R_{\rm f}])$ results
from optimizing over a larger set of parameters than $W_{\rm max}$. Note
that ${\rm tr}(H_{\rm S}[\rho_{\rm i}-\rho_{\rm f}])$ appears also in
the right-hand-side of (\ref{baratino}), and there is some analogy
between $W_{\rm sur}$ and the entropy difference $T({\rm tr}[- \rho_{\rm
eq}\ln\rho_{\rm eq} +\rho\ln\rho ])$ in (\ref{baratino}), which is the
work extracted from the canonical bath. There the work $W_{\rm max}$
extracted from the canonical equilibrium bath alone (without the
system) is zero.

Scalar product (\ref{oro}) is calculated straightforwardly; for
clarity we focus on the thermodynamic limit regime (\ref{koza}):
\BEA
\label{barak}
W_{\rm max}+W_{\rm sur}&=& \delta[\pi_1 F_2+(\pi_0-\pi_1)F_1], \\
{\rm tr}(H_{\rm S}\rho_{\rm f})&=&\epsilon [\pi_1 P_2+(\pi_0-\pi_1)P_1],
\label{hast}
\EEA
where $W_{\rm max}$ is given by (\ref{brabant}) and
we defined for $k=1,2$:
\begin{gather}
F_k\equiv k(\ell_{1k}-1)+v^{\ell_{1k}\,}\,\frac{1+v^\alpha}{1-v}\left[\frac{1}{1-v}+\frac{\alpha v^\alpha}{1+v^\alpha}\right]\nonumber\\
+\alpha\,{\rm sign}(\ell_{1k}-\ell_{2k})\left[  k-\frac{v^{\ell_{2k}} (1+ v^{1+\alpha})}{1-v} \right], \\
P_k\equiv\frac{v^{\ell_{1k}+\alpha}}{1-v}+{\rm sign}(\ell_{1k}-\ell_{2k})\left[k-\frac{v^{\ell_{2k}}(1+v^{1+\alpha})  }{1-v}  \right],\nonumber\\
\ell_{1k}\equiv\left\lceil \frac{\ln\left(\frac{k(1-v)}{1+v^\alpha}\right)}{\ln v}\right\rceil , \quad
\ell_{2k}\equiv\left\lceil \frac{\ln\left(\frac{k(1-v)}{1+v^{1+\alpha}}\right)}{\ln v}\right\rceil .
\label{bedlam}
\end{gather}
Recall that ${\rm sign}(0)=0$, $\alpha=\lfloor \frac{\epsilon}{\delta}\rfloor$, and that
$\lfloor x \rfloor$ and
$\lceil x \rceil$ are defined after (\ref{oro}) and (\ref{brabant}), respectively.

The final state $\rho_{\rm f}$ of the two-level system is diagonal in
its energy representation. The eigenvalues of $\rho_{\rm f}$ are
read-off from (\ref{hast}). The excited state of $\rho_{\rm f}$ is
less populated than the ground-state; otherwise it can still provide
work via a cyclic process. In this specific sense the two-level system
{\it partially} equilibrates; recall that $\rho_{\rm i}$ is arbitrary.

Fig.~\ref{fig1} displays $W_{\rm max}(T)$ and $W_{\rm sur}(T)$ for a
representative range of parameters.  It is seen that for $\alpha=0$,
$W_{\rm sur}\geq 0$ and both $W_{\rm max}$ and $W_{\rm sur}$
monotonically increase with $T$. For $\alpha>0$ the positivity of
$W_{\rm sur}$ is recovered only for a sufficiently high $T$ provided
that $\pi_0\not=\pi_1$; see Fig.~\ref{fig1}. For $\pi_0=\pi_1$, we
always get $W_{\rm sur}<0$.

Consider now $\epsilon<\delta$ [i.e. $\alpha=0$ in (\ref{oro})]
and assume $v\equiv e^{-\delta/T}\leq 1/3$ for simplicity.
Eqs.~(\ref{barak}--\ref{bedlam}) produce
\BEA
\label{ogust}
W_{\rm sur}&=&v\delta(2\pi_0-1)\,(1-v)^{-2},\\
\label{nansen}
{\rm tr}(H_{\rm S}\rho_{\rm f})&=&\epsilon[\,v\,\pi_0+(1-3v)\pi_1\,]\,(1-v)^{-1}.
\EEA
\comment{  
\BEA
\label{ogust}
W_{\rm sur}&=&\frac{a\delta(2\pi_0-1)}{(a-1)^2}~~{\rm for}~~a\equiv e^{\delta/T}\geq 3, \\
           &=&\frac{[a(a-3)+3]\delta(2\pi_0-1)}{(a-1)^2}~~{\rm for}~~3\geq a \geq 2, \nonumber\\
\label{nansen}
{\rm tr}(H_{\rm S}\rho_{\rm f})&=&\epsilon[\frac{\pi_0}{a-1}+\frac{a-3}{a-1}\pi_1] \,\,\,\,\,\,{\rm for}~~a \geq 3, \\
                               &=& \epsilon[\frac{a-2}{a-1}\pi_0+\frac{3-a}{a-1}\pi_1] \,\,{\rm for}~~3\geq a \geq 1+\sqrt{2}, \nonumber\\
                               &=& \epsilon[\frac{\pi_0}{a(a-1)}+\frac{\pi_1}{a}]\,\,\,\,\,\,{\rm for}~~1+\sqrt{2}\geq a \geq 2. \nonumber
\EEA}
Eq.~(\ref{ogust}) shows that, in addition to $W_{\rm sur}+{\rm
tr}(H_{\rm S}[\rho_{\rm i}-\rho_{\rm f}])\geq 0$, the work extracted
from the bath is enhanced, $W_{\rm sur}>0$, for any state of the
two-level system besides the completely mixed one, where
$\pi_0=\pi_1=\frac{1}{2}$.

The energy difference ${\rm tr}(H_{\rm S}[\rho_{\rm i}-\rho_{\rm f}])$
can be positive or negative. Hence parameters can be tuned such that it
is zero, e.g., from (\ref{nansen}) and for $\epsilon=T\ln 2<T\ln
3<\delta$ we get that initially canonical {\it equilibrium} two-level
system, $\pi_0=1-\pi_1=(1+e^{-\epsilon/T})^{-1}=\frac{2}{3}$, enhances the work
extracted from the bath {\it without} changing its marginal state: ${\rm
tr}(H_{\rm S}\rho_{\rm f})=0$.  Hence for enhancing the work extracted
from the microcanonical bath one needs that the system is
ordered: its state should not be completely mixed, while the maximal
enhancing is achieved for a pure state. But the state of the
system need not change.

A non-equilibrium system coupled to canonical equilibrium bath can
enhance the work extracted from the bath {\it only} at the cost of
changing (towards equilibrium) its initially non-equilibrium state; see
(\ref{baratino}). The free energy measures this change. For the
microcanonical bath, the work can be enhanced already by an equilibrium
two-state system without changing its marginal state. We conclude that
the concept of the free energy does not {\it generally} apply to a
system coupled to a microcanonical bath.

But this concept applies in the high-temperature limit.
For $T\gg\delta,\epsilon$ we get from (\ref{barak}--\ref{bedlam}) and from (\ref{gogor}):
\BEA
\label{oruk}
&&W_{\rm sur}=(1-2\pi_1)T\ln 2, \\
&&{\rm tr}(H_{\rm S}\rho_{\rm f})={\epsilon}/{2}.
\label{ongar}
\EEA
Eq.~(\ref{ongar}) means that the final state of the two-level system is
completely mixed, which for the present high-temperature case coincides
with the canonical equilibrium state. Eq.~(\ref{oruk}) predicts
work-enhancement only if initially the two-level system was out of
equilibrium [recall that $\frac{1}{2}\geq\pi_1$]. Hence for $T\gg\delta,\epsilon$
we recover the logics of the canonical-bath situation, but not its letter,
because for a canonical bath $W_{\rm sur}$ reduces to the difference
between two von Neumann entropies that are logarithmic functions of the
initial eigenvalues $\pi_0$ and $\pi_1$; cf. our remark after (\ref{gal}).

We shall show elsewhere that a for a $\mu$-level system in a state (density matrix)
$\rho$ with eigenvalues ordered as $\pi_0\geq\pi_1\geq...\geq\pi_{\mu-1}$,
we can define the {\it linear entropy} as
\BEA
\label{lin}
{\cal L}[\rho] = {\sum}_{k=1}^{\mu-1}\, \pi_k [(k+1)\ln(k+1)-k\ln k].
\EEA
Generalizing (\ref{oruk}), the surplus work $W_{\rm sur}$ extracted from a high-temperature
microcanonical bath in contact with this system is then $W_{\rm
sur}=T(\ln\mu-{\cal L}[\rho])=T({\cal L}[\frac{\hat{1}}{\mu}]-{\cal
L}[\rho])$, where $\frac{\hat{1}}{\mu}$ is the maximally mixed state of
the $\mu$-level system. For the canonical situation this expression
involves the von Neumann entropy $-{\rm tr}[\rho\ln\rho]$ instead of
${\cal L}[\rho]$.  Note that $\ln\mu\geq {\cal L}[\rho]\geq 0$: the
upper (lower) limit is reached for the maximally mixed (pure) $\rho$.

Hence for a system in initial state $\rho$ and Hamiltonian $H_{\rm S}$
coupled to the {\it high-temperature} microcanonical bath one can define
the {\it microcanonical} free energy ${\cal F}[\rho]={\rm tr}(H_{\rm
S}\rho)-T{\cal L}[\rho]$, whose difference ${\cal F}[\rho]-{\cal
F}[\rho_{\rm eq}]$ (after adding to $W_{\rm max}$ extracted from the
bath alone) defines the maximal work extracted from the
system+bath.

{\bf Summary}. We reformulated the Thomson's formulation of the second
law for a $N\gg 1$ particle {\it equilibrium} bath in a microcanonical
state: if the bath temperature $T$ is finite, the maximal work
extractable from the bath via a cyclic Hamiltonian process is $\gtrsim
\delta$, where $\delta$ is the energy of a single bath particle.  The
maximal work tends to $\delta\sqrt{N\ln N}$ if $N$ is large but fixed
and $T\to \infty$. The reformulation applies equally well to both
ordinary microcanonical states, which are macroscopically equivalent to
canonical states, and convex-entropy microcanonical states for which no
canonical state can be defined, e.g., because of a negative specific
heat \cite{longrange}. The existence of such states demonstrates that a
viewpoint on a microcanonical state as emerging from measuring the
energy of the canonical state is not generally valid.  Thermodynamics of
such systems can have peculiarities \cite{leyvraz}, but we saw that they
satisfy the same generalized Thomson's formulation much in the same way
as ordinary microcanonical states. The work extraction is possible,
since the microcanonic state of the bath is not Gibbsian, though
each its constituent can be in a Gibbsian state.

It is widely known that only a non-equilibrium system can lead|at expense of
changing its state towards equilibration|to work extraction from a
canonical bath \cite{landau}. This work is given by the free energy
difference (\ref{baratino}). In contrast, a canonical equilibrium system
(having the same temperature as the bath) can enhance the work extracted
from the microcanonical bath without changing its marginal state. Hence
the concept of free energy, in the sense of the maximal work, does not
generally apply to the microcanonical situation. The application of the
concept is recovered for $T\gg \delta$, but the canonical expression of
the free energy is not restored, instead it should be formulated via the
linear entropy (\ref{lin}).

\acknowledgements{
We thank G. Mahler, D. Janzing and R. Balian for
discussions. We were supported by Volkswagenstiftung. }

\comment{We thank our colleagues Guenter Mahler, Dominik Janzing and Roger Balian for
discussions that motivated this work. We were supported by Volkswagenstiftung. }


\begin{thebibliography}{9}
\bibitem{f0}The maximization of work is carried out
over all cyclic and time-dependent (in the sense of (\ref{dumbadze}))
Hamiltonians acting on the system+bath \cite{landau}; for more details
see, e.g., Appendix E of \cite{johal}. One can show that the work
(\ref{baratino}) can be achieved for a Hamiltonian, where external fields
act only on the system and on its coupling to the bath \cite{polish,johal}.
This set-up is local (advantage), but it has to be slow (drawback).
\bibitem{f1} Eq.~(\ref{entropy}) is one of two standard definitions of
the microcanonic entropy \cite{landau}. Another definition is
$\ln {\rm tr}[\rho P_{E,E_0}]$, where $E_0$ is the
lowest energy and $P_{E,E_0}$ is the projector on
the subspace $[E,E_0]$. For macroscopic systems both definitions typically agree with each other
within the order ${\cal O}(N)$, where $N\gg 1$ is the number of particles. For finite
systems the second definition is preferable \cite{berdi,armen_epl}.
\bibitem{f2}It is derived from (\ref{kaban}) via the Stirling's formula
$N!\simeq (N/e)^N$ and expanding over $m$.
\end{thebibliography}
\end{document}